\documentclass{cimento}
\usepackage{graphicx}
\newcommand{\be}{\begin{equation}}
\newcommand{\ee}{\end{equation}}
\newcommand{\bea}{\begin{eqnarray}}
\newcommand{\eea}{\end{eqnarray}}
\newcommand{\bean}{\begin{eqnarray*}}
\newcommand{\eean}{\end{eqnarray*}}
\title{The spinor field in Rindler spacetime: \\ an analysis of the Unruh effect}
\author{D.~Oriti\from{ins:icra}\thanks{on moving to: Department of Applied Mathematis and Theoretical Physics, University of Cambridge, Silver Street, Cambridge CB3 9EW, UK}} 
\instlist{\inst{ins:icra} I.C.R.A. - International Center for Relativistic Astrophysics, \\ Dipartimento di Fisica, University of Rome "la Sapienza", P.le A. Moro, 5, 00185, Rome - Italy \\ oriti@icra.it}
\begin{document}
\maketitle
\begin{abstract}
We analyse the quantization procedure of the spinor field in the Rindler spacetime, showing the boundary conditions that should be imposed to the field, in order to have a well posed theory. We then investigate the relationship between this construction and the usual one in Minkowski spacetime. This leads to the concept of "Unruh effect", that is the thermal nature of the Minkowski vacuum state from the point of view of an accelerated observer. It is demostrated that the two constructions are qualitatively different and can not be compared and consequently the conventional interpretation of the Unruh effect is questionable.
\end{abstract}
\section{Introduction}
It is a widely known result that one obtained by Unruh ~\cite{Unruh} almost 25 years ago, and now named after him as \lq\lq Unruh effect". Resuming it could be expressed in the following statements: 
\begin{itemize}
\item the Minkowski vacuum state, from the point of view of an accelerated observer, is a particle state described by a density matrix at the temperature
\be T\,=\,\frac{a}{2\,\pi\,k_{B}} \label{eq:T} \ee called the Unruh-Davies temperature,
where $a$ is the (constant) acceleration of the observer and $k_{B}$ is the Boltzmann constant;
\item an accelerated observer in the empty Minkowski space will detect a thermal bath of particles at the temperature (~\ref{eq:T}).
\end{itemize}
It should be noted that two problems are involved here, which in principle are different, but which are claimed to be equivalent: 
\begin{itemize}
\item the physical properties of a quantum field when restricted to a submanifold (the Rindler Space, (RS)) of Minkowski space (MS);
\item the behaviour of a constantly accelerated particle detector in empty flat space.
\end{itemize} 
The first one deals only with basic principles of quantum theory and appears to be in a sense more fundamental, whether the second one involves also a description of structure and characteristics of the detector, and details of interaction with the quantum field. Moreover, it is the first analysis that made possible to claim for universality of the Unruh-like detector response. We will treat here only the first problem above, because of its importance in understanding what happens when one tries to analyse a field in a submanifold of a maximally analytically extended manifold, and also because it seems to be related in many ways with the phenomenon of quantum evaporation of black holes, i.e. the Hawking effect ~\cite{Unruh}~\cite{Hawking}~\cite{DesLev}~\cite{CadoMign}~\cite{Wald}~\cite{BirDav}.

The procedure used by Unruh is based on a quantization scheme for a free field in MS, alternative but claimed to be equivalent to the standard one, which uses as the Hilbert space of solutions of the wave equation  
\be \mathcal{H}_{U}\,=\,\mathcal{H}_{R}\,\oplus\,\mathcal{H}_{L} \ee where $\mathcal{H}_{R}$ consists of solutions which are non-zero everywhere but in the L sector, which have positive frequency with respect to the \lq\lq Rindler time" $\eta$, and which reduce themselves to the well known Fulling modes ~\cite{Fulling}, and $\mathcal{H}_{L}$ is given by the solutions which are non-zero everywhere but in R and with negative frequency with respect to $\eta$. Then it is obtained a representation of the Minkowski vacuum state as a state in $\mathcal{F}_{s}(\mathcal{H}_{U})$, i.e. the Fock space constructed on $\mathcal{H}_{U}$. Finally it is derived the particle content of Minkowski vacuum and the expression for the density matrix associated to it, when expressed as a mixed state in the Fock space $\mathcal{F}_{s}(\mathcal{H}_{R})$ having traced out the degrees of freedom related to the L sector, which is unaccessible to the Rindler observer.

The usual explanation of the Unruh effect is based exactly on the presence, for a Rindler observer, which is confined inside the Rindler wedge, of an event horizon which prevent him from having part of the informations about the quantum field, so that he sees Minkowski vacuum state as a mixed state. But this explanation is in our opinion of little utilily for two main reason: first, the existence of horizons is due to overidealization of the problem, since for physical accelerations (which last finite amount of time) no horizon should be present, so that the response of an accelerated detector, if of the Unruh-Davies type, cannot be caused by them; second, in the quantum theoretical treatment of the problem, the presence of event horizons is aspected to affect deeply the fields, from the point of view of Rindler observer, in the form of some kind of boundary condition, which instead are totally absent in the Unruh scheme.

What we are going to do in this paper is to show the conditions to (and only to) which the quantum theory of the spinor field in the Rindler spacetime is well posed, to analyse the relationship between this construction and the usual one in MS, in order to find which role is played by these conditions in the derivation and interpretation of the Unruh effect in the spinor case, and to understand finally what is its physical significance.

What we will find is that a correct quantization procedure for the spinor field in Rinlder space requires the boundary condition
\be \lim_{\rho\,\rightarrow\,0}\,\rho^{\frac{1}{2}}\,\Psi\left( 0\,,\,\rho\right)\,=\,0 \label{eq:cond} \ee

i.e. the field should not grow up too rapidly at the origin of Minkowski (and Rindler) space. This means that the quantization on a background manifold which is not maximally extended requires a boundary condition which is absent in the quantization procedure over the extended manifold; in the literature this is not recognized clearly enough, and the usual procedure is to restrict the fields just considering a smaller domain of definition.

Moreover, we show that the same boundary condition is implied in the standard derivation of the Unruh effect, i.e. in the comparison between the quantum construction in RS and in MS. On the one hand this boundary condition represents a key point in the usual derivation, which was not considered before, but on the other hand it severely undermines the significance of the Unruh effect itself, and permits to assert that the traditional interpretation of it is inconsistent and that the basic principles of Quantum Field Theory give no grounds to the existence of an Unruh effect.

These represent a generalization to the spinor field of similar results recently obtained for the scalar field ~\cite{Bel1}~\cite{Bel2}~\cite{Bel3}; consequently, they appear to be consequences of very general properties of Quantum Field Theory, and seem to be firmly established.

In addition to the main results cited, we obtained some minor but original results, necessary to achieve the first, namely: the explicit expression of the Lorentz boost generator (or Lorentz Momentum) for the spinor field, its eigenfunctions and their analytical representation holding in the whole MS (except for the origin), which is in turn a generalization of the Gerlach's Minkowski Bessel Modes ~\cite{Gerlach}.
    

\section{Rindler Spacetime} \label{sec:Rind}
Let's consider a particle moving in MS with constant acceleration $a$ along the x-axis; it will follow the trajectory given by (parameter $\tau$):
\be t\,=\,a^{-1}\sinh(a\tau)\;\;\;\;\;x\,=\,a^{-1}\cosh(a\tau)\;\;\;y\,=\,y(0)\;\;\;z\,=\,z(0) \label{eq:acc} \ee
This is an hyperbola in the (t,x) plane. The lines $t=\pm x$ represent asymptotes for it and event horizons for the moving particle. Varying $a$ we obtain different hyperbolas with the same characteristics. Let's now perform, starting from the Minkowski metric, the change in coordinates given by:
\bea  t\,=\,\rho\,\sinh\,\eta\;\;\;\;\;\;x\,=\,\rho\,\cosh\,\eta \label{eq:cooRin} \label{eq:coo}\\
 \rho\,=\,\sqrt{x^{2}-t^{2}}\;\;\;\;\;\;\eta\,=\,arctgh\left(\frac{t}{x}\right) \eea          
with the other coordinates left unchanged.

The metric assumes the form:
\be ds^{2}\,=\,\rho^{2}d\eta^{2}\,-\,d\rho^{2}\,-\,dy^{2}\,-\,dz^{2}  \label{Rindler} \ee
Note that it describes a stationary spacetime. The worldlines $\rho=const, y=const, z=const$ correspond to uniformly accelerated observers, with $a=\rho^{-1}$ and proper time $\tau=\rho\eta$, as it can be seen comparing (\ref{eq:acc}) and (\ref{eq:coo}). We can think at the Rindler Space as the collection of these worldlines, and this is the reason why RS is generally regarded as the \lq\lq natural\rq\rq manifold in which to describe accelerate motion. The hypersurfaces $\eta=const$ describe events which are simultaneous from the point of view of a \lq\lq Rindler (uniformly accelerated) observer". The Rindler manifold cannot be extended to negative values of $\rho$ trough $\rho=0$, for RS is no more rigid beyond the limit $\rho\rightarrow 0$. This hypersurface, in fact, represents, as we said, an event horizon for Rindler observers, which cannot see any event located beyond it. Nevertheless the horizon is a regular surface, of course, and the metric singularity is due only to the choice of the coordinates. 

\includegraphics[width=8cm]{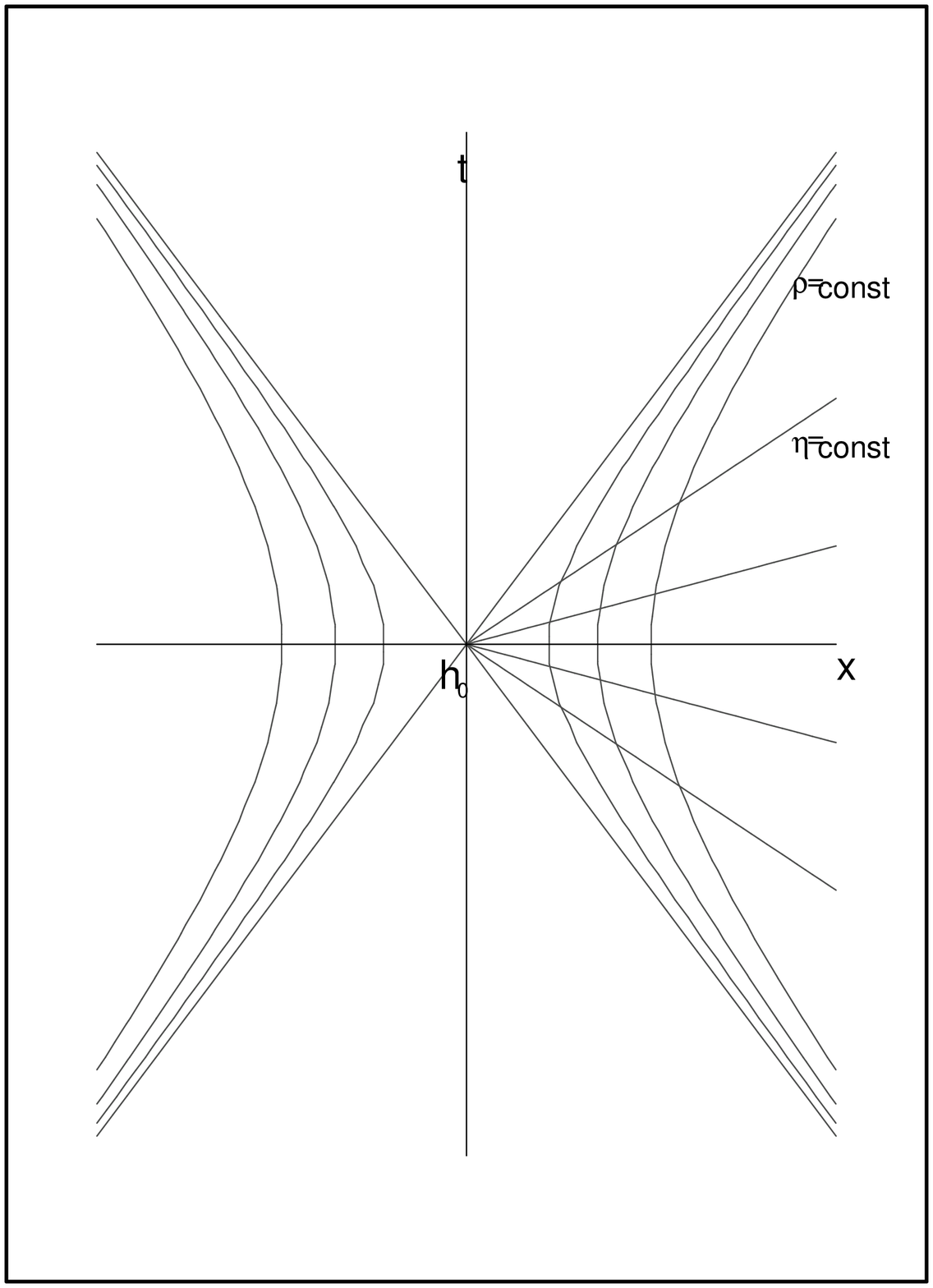}

The (~\ref{eq:cooRin}) cover only a sector of the whole MS and the others are covered by the following charts:
\begin{center}
{L\,\,:}\be t\,=\,\rho\,\sinh\,\eta\;\;\;x\,=\,\rho\,\cosh\,\eta \ee
\be \rho\,=\,-\,\sqrt{x^{2}-t^{2}}\;\;\;\eta\,=\,arctgh(\frac{t}{x}) \ee
{F\,\,:}
\be t\,=\,\rho\,\cosh\,\eta\;\;\;x\,=\,\rho\,\sinh\,\eta \ee
\be \rho\,=\,\sqrt{t^{2}-x^{2}}\;\;\;\eta\,=\,arctgh(\frac{x}{t}) \ee
{P\,\,:}
\be t\,=\,\rho\,\cosh\,\eta\;\;\;x\,=\,\rho\,\sinh\,\eta \ee
\be \rho\,=\,-\,\sqrt{t^{2}-x^{2}}\;\;\;\eta\,=\,arctgh(\frac{x}{t}) \ee
\end{center}
It's important to note also that these characteristics of RS inside MS are analogous to that of Schwarzchild spacetime inside the Kruskal one, from both geometrical and physical point of view.
         
\section{Quantization in Rindler space} \label{sec:RQFT}
We now turn to the problem of quantization of the spinor field in the Rindler wedge. The procedure will be the standard one, so we will first solve the Dirac equation looking for hamiltonian eigenfunctions.
\subsection{The Dirac equation and its solutions in RS}
Using the tetrad formalism, the Dirac equation in a generic curved spacetime is given by ~\cite{BrilWhe}:
\be (\,i\,\gamma^{\mu}\,\nabla_{\mu}\,-\,m\,)\,\Psi\,=\,0 \label{eq:D} \ee
where:
\begin{itemize}
\item  $\gamma^{\mu}\,=\,\theta^{\mu}_{\bar{\mu}}\,\gamma^{\bar{\mu}}$ are the analogous in curved spaces of the usual Dirac gamma matrices $\gamma^{\bar{\mu}}$, and they satisfy:
\be \gamma^{\mu}\,\gamma^{\nu}\,+\,\gamma^{\nu}\,\gamma^{\mu}\,=\,g^{\mu\nu} \ee
\item la  $\theta^{\mu}_{\bar{\mu}}$ is the inverse of the tetrad vector $\theta^{\bar{\mu}}_{\mu}$ with vectorial index $\mu$ and tetradic index $\bar{\mu}$;
\item $\nabla_{\mu}=\partial_{\mu}-\Gamma_{\mu}$ is the spinorial covariant derivative, defined in such a way that $\nabla_{\mu}\Psi$ is a covariant vector;
\item \be \Gamma_{\mu}\,=\,\frac{1}{8}\,\theta_{\bar{b}}^{b}\,\theta_{\bar{c}\,b\,;\,\mu}\,[\gamma^{\bar{c}},\gamma^{\bar{b}}] \ee is the spinorial connection.
\end{itemize}

Let's consider the particular case of the Rindler metric:
\be ds^{2}\,=\,\rho^{2}\,d\eta^{2}\,-\,d\rho^{2}\,-\,dy^{2}\,-\,dz^{2} \label{eq:metrica} \ee
(From this point we will omit the bar over the gamma matrices indexes).

The Dirac equation 
assumes the form:
\bea \left(  i\,\partial_{\eta}+i\,\rho\,\gamma^{0}\gamma^{1}\partial_{\rho}+i\,\rho\,\gamma^{0}\gamma^{2}\partial_{y}+i\,\rho\,\gamma^{0}\gamma^{3}\partial_{z}+\frac{i}{2}\gamma^{0}\gamma^{1}-m\,\rho\,\gamma^{0} \right) \Psi=0 \\ \Longrightarrow \;\;\;\;\;\;\;\;i\,\partial_{\eta}\,\Psi\,=\left(\,-\,i\,\rho\,\alpha_{i}\,\partial_{i}\,-\,\frac{1}{2}\,i\,\alpha_{1}\,+\,m\,\rho\,\beta \right) \Psi \eea
that is a Shroedinger-like form with an hamiltonian given by:
\bea H_{R}\,=\,-\,i\,\rho\,\alpha_{i}\,\partial_{i}\,-\,\frac{1}{2}\,i\,\alpha_{1}\,+\,m\,\rho\,\beta \,= \;\;\;\;\;\;\;\;\;\;\;\;\;\;\;\;\;\;\;\;\;\;\;\;\;\;\;\;\;\;\;\;\\ =\,-\,i\,\rho\,\gamma^{0}\,\gamma^{1}\,\partial_{\rho}\,-\,i\,\rho\,\gamma^{0}\,\gamma^{2}\,\partial_{y}\,-\,i\,\rho\,\gamma^{0}\,\gamma^{3}\,\partial_{z}\,-\,\frac{1}{2}\,i\,\gamma^{0}\,\gamma^{1}\,+\,m\,\rho\gamma^{0} \label{eq:H} \eea
We now look for solutions of the Dirac equation which are simultaneously eigenfunctions of the Rindler hamiltonian (~\ref{eq:H}) and of the operators $P_{y}$ and $P_{z}$ (of course they should also have the correct behaviour at infinity). So we expect to find a degeneracy of these solutions, because we know that three operators are not sufficient to completely characterize the states of the spinor field. This degeneracy is however of no relevance for our pourposes, so we will not deal with it.

The solutions we look for are the following:
\be \Psi_{1\mathcal{M}}^{R}\,=\,N_{\mathcal{M}}\,\left( X_{1}^{R}\,K_{i\,\mathcal{M}-\frac{1}{2}}(\kappa\rho)\,+\,Y^{R}_{1}\,K_{i\,\mathcal{M}+\frac{1}{2}}(\kappa\rho) \right)\,e^{-\,i\,\mathcal{M}\,\eta}\,e^{i\,k_{2}\,y\,+\,i\,k_{3}\,z}\label{eq:Psi1} \ee

with: 

\be X^{R}_{1}\,=\,\left( \begin{array}{c} k_{3} \\ i\,\left(k_{2}\,+\,i\,m \right) \\ i\,\left(k_{2}\,+\,i\,m \right) \\ k_{3} \end{array} \right) \;\;\;\;\;\;\;\;\;\;\;\;\;\;Y^{R}_{1}\,=\,\left( \begin{array}{c} 0 \\ i\,\kappa \\ -\,i\,\kappa \\ 0 \end{array} \right) \label{eq:XY1} \ee

and

\be \Psi^{R}_{2\mathcal{M}}\,=\,N_{\mathcal{M}}'\,\left( X^{R}_{2}\,K_{i\,\mathcal{M}-\frac{1}{2}}(\kappa\rho)\,+\,Y^{R}_{2}\,K_{i\,\mathcal{M}+\frac{1}{2}}(\kappa\rho) \right)\,e^{-\,i\,\mathcal{M}\,\eta}\,e^{i\,k_{2}\,y\,+\,i\,k_{3}\,z} \label{eq:Psi2} \ee

with:

\be X^{R}_{2}\,=\,\left( \begin{array}{c} 0 \\ i\,\kappa \\ i\,\kappa \\ 0 \end{array} \right) \;\;\;\;\;\;\;\;\;\;\;\;\;\; Y^{R}_{2}\,=\,\left( \begin{array}{c} k_{3} \\ i\,\left(k_{2}\,-\,i\,m \right) \\ -\,i\,\left(k_{2}\,-\,i\,m \right) \\ -\,k_{3} \end{array} \right) \label{eq:XY2} \ee
 
where $k_{2}$,$k_{3}$ are eigenvalues of $P_{2}$, $P_{3}$, $\kappa=\sqrt{k_{2}\,+\,k_{3}\,+\,m^{2}}$, $\mathcal{M}$ is the eigenvalue of the hamiltonian, and $N_{\mathcal{M}}$, $N'_{\mathcal{M}}$ are normalization factors which are found to be:
\be N_{\mathcal{M}}\,=\,N_{\mathcal{M}}'\,=\,\frac{1}{4\,\pi^{2}\,\sqrt{\kappa}}\,\sqrt{\cosh{\pi\,\mathcal{M}}} \ee
\subsection{Physical characterization of the solutions}
We now want to understand better the physical nature of the solutions (~\ref{eq:Psi1})(~\ref{eq:Psi2}), and this means to characterize them in a clearer way than just saying they are eigenfunctions of the Rindler hamiltonian.

For this pourpose we need to find the expressions for the solutions in minkowski coordinates.

In order to do it, it is necessary to consider carefully the way in which spinors transform under coordinate transformations. It is well known that in the formalism we have used, spinors are scalars, or, more precisely, a collection of four scalars, and this is what makes the Dirac  equation (~\ref{eq:D}) covariant. But after a coordinate transformation, also the tetrads undergo to a transformation, namely an orthogonal rotation, since the observer associated with the new coordinate system will build up a new tetrad, which is in principle different from the one used in the previous coordinate system (the transformation should of course be orthogonal so that the new tetrad vectors satisfy their definitory equation). This orthogonal rotation of the tetrad will determine a transformation of the spinor field, and the correspondent operator is to be found analysing the coordinate transformation itself, linearizing it (we are looking for orthogonal, i.e. linear, transformation) and working in the local minkowskian neighbour of the point we are working in. The result is that the spinor transformation corresponding to the Rindler coordinate transformation (~\ref{eq:cooRin}) is given by:
\be \Psi(t\,,\,x)\,=\,S(\eta)\,\Psi(\eta\,,\,\rho)\,=\,\exp{\left( \frac{1}{2}\,\gamma^{0}\,\gamma^{1}\,\eta\right)}\,\Psi(\eta\,,\,\rho) \ee

Note that the operator we found, $S=\exp{\left( \frac{1}{2}\gamma^{0}\gamma^{1}\eta\right)}$, has the form of an operator resulting from a Lorentz coordinate transformation, with "velocity" parameter $\eta$.

The solutions (~\ref{eq:Psi1})-(~\ref{eq:Psi2}) in minkowskian coordinates take the form: 
\be \Psi_{i,\mathcal{M}}^{R}(t,x,y,z)\,=\,N_{\mathcal{M}}\,\left( X^{R}_{i}\,K_{i\,\mathcal{M}-\frac{1}{2}}(\kappa\rho)\,e^{-\,\left(i\,\mathcal{M}-\frac{1}{2}\right)\eta}\,+\,Y^{R}_{i}\,K_{i\,\mathcal{M}+\frac{1}{2}}(\kappa\rho)\,e^{-\,\left(i\,\mathcal{M}+\frac{1}{2}\right)\eta} \right)\,e^{i\,k_{2}\,y\,+\,i\,k_{3}\,z} \ee

It must be clear that these are always defined only in the RS, but just expressed in minkowskian coordinates, and this is why the normalization factor is still $N$.

Now we can turn to the anticipated physical characterization of the solutions: they are found to be eigenfunctions of the Boost Generator Operator, or Lorentz Momentum. This could be aspected, because 1)$\Psi_{i\mathcal{M}}$ were eigenfunctions of the Rindler Hamiltonian $H_{R}$ and
2)$H_{R}$ is precisely the Lorentz Boost Generator written in Rindler coordinates, since
3)the time evolution of a Rindler (uniformly accelerated) observer is properly a infinite succession of infiniteimal boost transformations.

It is easy to verify this statement, once known the Lorentz Momentum operator, and this in turn can be obtained from the classical theory of fields.

In fact, given the conserved quantity:
\be M^{0 i}\,=\,\int\,d^{3}x\,\left[ \left( x^{0}\,T^{i 0}\,-\,x^{i}\,T^{0 0} \right) \,+\,S^{0 i 0} \right] \ee
corresponding to the invariance of the Lagrangian under boost transformations along the i-axis, (which are isometries of the Rindler spacetime), the explicit calculation of $T^{\mu\nu}$ and of $S^{\mu\nu}$ gives:
\be M^{01}\,=\,\int\,d^{3}x\,\Psi^{\dagger}\,\left[ i\,t\,\partial^{i}\,+\,x^{i}\,\left( i\,\gamma^{0} \,\gamma^{i} \,\partial_{i} \,-\,m\,\gamma^{0} \right) +\, \frac{i}{2}\,\gamma^{0}\,\gamma^{i}\right]\,\Psi \label{eq:Mcl} \ee
Interpreting it as a mean value of a quantum operator, we obtain for the Lorentz Momentum the expression (for contravariant and covariant components):
\be  M^{0 i}\,=\,-\,i\,t\,\partial_{i}\,-\,x_{i}\,\left( -\,i\,\gamma^{0}\,\gamma^{j}\,\partial_{j}\,+\,m\,\gamma^{0} \right) +\,\frac{i}{2}\,\gamma^{0}\,\gamma^{i} \ee
 \be  M_{0 i}\,=\,+\,i\,t\,\partial_{i}\,+\,x_{i}\,\left( -\,i\,\gamma^{0}\,\gamma^{j}\,\partial_{j}\,+\,m\,\gamma^{0} \right) -\,\frac{i}{2}\,\gamma^{0}\,\gamma^{i} \label{eq:M} \ee
where the symbols $\partial_{j}$ are intended to represent just partial derivatives, with no particular covariant characterization.

Given this expression it can be verified that our solutions are eigenfunctions of the operator $M_{01}$ with eigenvalue $\mathcal{M}$, and this represent their physical characterization.
\subsection{The second quantization and its conditions}
We now possess all the necessary elements to perform the second quantization of the spinor field in RS, i.e. a set of normalized functions which are solutions of the equation of motion.

We then expand the field in terms of them:
\bea \lefteqn{\Psi^{R}(\eta,\rho,y,z)\,=\,\sum_{i=1,2}\,\int_{-\infty}^{+\infty}d\mathcal{M}\,\int_{-\infty}^{+\infty}dk_{2}\,\int_{-\infty}^{+\infty}dk_{3}\,a_{\mathcal{M}i}(k_{2},k_{3})\,\Psi_{i,\mathcal{M},k_{2},k_{3}}(t,x,y,z)\,=} \nonumber \\ &=& \sum_{i=1,2}\,\int_{0}^{+\infty}d\mathcal{M}\,\int_{-\infty}^{+\infty}dk_{2}\,\int_{-\infty}^{+\infty}dk_{3}\,\times \nonumber \\ & \times& \left( a_{\mathcal{M},i}(k_{2},k_{3})\,\Psi^{R}_{i,\mathcal{M},k_{2},k_{3}}(\eta,\rho,y,z)\,+\,b^{\dagger}_{\mathcal{M},i}(k_{2},k_{3})\,\Psi^{R}_{i,-\mathcal{M},k_{2},k_{3}}(\eta,\rho,y,z)\right)  \label{eq:exp} \eea
The second quantization is now performed considering the coefficients $a_{i,\mathcal{M}}$ and $b_{i,\mathcal{M}}$ (and their hermitian coniugated) as operators, and requiring for their anticommutators:
\bea \left\{ a_{\mathcal{M},i}(k_{2},k_{3})\,,\,a_{\mathcal{M}',j}^{\dagger}(k'_{2},k'_{3})\right\}\,=\,\delta_{ij}\,\delta(\mathcal{M}-\mathcal{M}')\,\delta(k_{2}-k'_{2})\,\delta(k_{3}-k'_{3})\;\;\; \\ \left\{ b_{\mathcal{M},i}(k_{2},k_{3})\,,\,b_{\mathcal{M}',j}^{\dagger}(k'_{2},k'_{3})\right\}\,=\,\delta_{ij}\,\delta(\mathcal{M}-\mathcal{M}')\,\delta(k_{2}-k'_{2})\,\delta(k_{3}-k'_{3})\;\;\;  \\ 
\left\{ a_{\mathcal{M},i}(k_{2},k_{3})\,,\,b_{\mathcal{M}',j}^{\dagger}(k'_{2},k'_{3})\right\}\,=\,0\;\;\;\;\forall\;i,j,\mathcal{M},\mathcal{M}',k_{2},k_{2}',k_{3},k_{3}'\;\;\;\;\;\;\;\;\;\;\;\; \\ \left\{ a_{\mathcal{M},i}(k_{2},k_{3})\,,\,b_{\mathcal{M}',j}(k'_{2},k'_{3})\right\}\,=\,0\;\;\;\;\forall\;i,j,\mathcal{M},\mathcal{M}',k_{2},k_{2}',k_{3},k_{3}'\;\;\;\;\;\;\;\;\;\;\;\; \eea 

and the quantum states of the field are constructed from the Rindler vacuum state $\mid 0\,\rangle_{R}$, defined by:
\be a_{\mathcal{M},i}(k_{2},k_{3})\,\mid 0\,\rangle_{R}\,=\,0\;\;\;\; \forall\;\;i,\,\mathcal{M},\,k_{2},\,k_{3} \ee

Now we will study if this quantum construction is well posed and at which conditions.

Let's consider again the Rindler hamiltonian:
\bea \lefteqn{H_{R}\,=\,-\,i\,\rho\,\alpha_{i}\,\partial_{i}\,-\,\frac{1}{2}\,i\,\alpha_{1}\,+\,m\,\rho\,\beta \,=} \\ &=&\,-\,i\,\rho\,\gamma^{0}\,\gamma^{1}\,\partial_{\rho}\,-\,i\,\rho\,\gamma^{0}\,\gamma^{2}\,\partial_{y}\,-\,i\,\rho\,\gamma^{0}\,\gamma^{3}\,\partial_{z}\,-\,\frac{1}{2}\,i\,\gamma^{0}\,\gamma^{1}\,+\,m\,\rho\gamma^{0}\;\;\;\;\;\;\;\;\;\;\;  \eea
and let's check whether it represents an hermitian operator, that is a necessary and sufficient condition for the completeness and orthonormality of the modes used. We should verify the condition
\be \left( H_{R}\Phi\,,\,\Psi\right)\,=\,\left( \Phi\,,\,H_{R}\Psi\right) \ee
with scalar product given by:
\be \left( \Phi\,,\,\Psi\right)\,=\,\int_{-\infty}^{+\infty}\,dy\,\int_{-\infty}^{+\infty}\,dz\,\int_{0}^{+\infty}d\rho\,\Phi^{\dagger}\,\Psi \ee
The explicit calculation shows that:
\be \left( H_{R}\Phi\,,\,\Psi\right)\,=\,\left( \Phi\,,\,H_{R}\,\Psi\right)\,+\,\int_{-\infty}^{+\infty}\,dy\,\int_{-\infty}^{+\infty}\,dz\,\left[ i\,\Phi^{\dagger}\,\alpha_{1}\,\rho\,\Psi \right]^{\rho\,=\,+\infty}_{\rho\,=\,0} \ee 
Then it is evident that the hermiticity of the hamiltonian is assured if and only if 
 \be \lim_{\rho\rightarrow\,0}\,\rho^{\frac{1}{2}}\,\Psi\left(\eta\,,\,\rho\,,\,y\,,\,z\right)\,=\,0 \;\;\;\;\forall\eta \label{eq:Condizione} \ee
and of course with analogous condition at $\rho\rightarrow +\infty$.

We emphasize that, since the field $\Psi$ is to be considered as an operator-valued distribution, this condition should be interpreted in the weak sense.

We stress again that a similar condition was already found for the scalar field in RS ~\cite{Bel1}~\cite{Bel2}~\cite{Bel3}. 

If the condition (~\ref{eq:Condizione}) is necessary for the hermiticity of the hamiltonian, it is expected to appear also in the analysis of the coefficients of the expansion (~\ref{eq:exp}). So we write the explicit expression of the coefficients $a_{i,\mathcal{M}}(k_{2},k_{3})$:
\bea \lefteqn{a_{\mathcal{M},i}(k_{2},k_{3})\,=\,\left( \Psi_{i,\mathcal{M},k_{2},k_{3}}\,,\,\Psi \right)_{R}\,=} \\ &=& +\,\int_{-\infty}^{+\infty}\,dy\,\int_{-\infty}^{+\infty}\,dz\,\int_{0}^{+\infty}d\rho\,N_{\mathcal{M}}^{*}\left[ \left( X_{i}^{\dagger}\,K_{i\mathcal{M}+\frac{1}{2}}\,+\,Y_{i}^{\dagger}\,K_{i\mathcal{M}-\frac{1}{2}}\right)\,\Psi\right]\,\times \nonumber \\ &\times& e^{(i\,\mathcal{M}\,\eta\,+\,i\,k_{2}\,y\,+\,i\,k_{3}\,z)} \label{eq:coeff} \eea

and consider the behaviour of the $K_{i\mathcal{M}\pm\frac{1}{2}}(\kappa\rho)$ for $\rho\simeq 0$. 
We have:
\bea \lefteqn{K_{i\mathcal{M}\pm\frac{1}{2}}(\kappa\rho)\,\simeq\,\frac{\pi}{2}\frac{1}{\sinh\left(\pi\left( i\mathcal{M}\pm\frac{1}{2}\right)\right)}\times} \nonumber \\ &\times&\left[ \frac{\kappa^{-\left( i\mathcal{M}\pm\frac{1}{2}\right)}}{\Gamma\left( -\left( i\mathcal{M}\pm\frac{1}{2}\right)\,+\,1\right)}\left( \frac{\rho}{2}\right)^{-\left( i\mathcal{M}\pm\frac{1}{2}\right)}\,-\,\frac{\kappa^{\left( i\mathcal{M}\pm\frac{1}{2}\right)}}{\Gamma\left( \left( i\mathcal{M}\pm\frac{1}{2}\right)\,+\,1\right)}\left( \frac{\rho}{2}\right)^{\left( i\mathcal{M}\pm\frac{1}{2}\right)}\right]\;\;\;\;\;\;\;\,\;\;\;\;\;\; \eea
We see that a divergence is present for $\rho\rightarrow 0$. Moreover we note that, for $\mathcal{M}=0$, we have the exact expression:
\be K_{\frac{1}{2}}(\kappa\rho)\,=\,\sqrt{\frac{\pi}{2\,\kappa\,\rho}}\,e^{-\,\kappa\,\rho} \ee
again with the same type of divergence.

It is so evident that, in order the integral (~\ref{eq:coeff}) to converge, is necessary to require the boundary condition (~\ref{eq:Condizione}) on the field. Otherwise, the coefficients $a_{i,\mathcal{M}}(k_{2},k_{3})$ are not defined and so it is for fundamental operators of quantm field theory like energy or particle number.

\subsection{Discussion}
Let us now discuss the result just found: the quantum theory of the spinor field in Rindler space is well defined if and only if we have the condition (~\ref{eq:Condizione}). This condition means that the field is to be quantized in a different way in MS and RS, because that the horizons are not only of a causal but also of a physical significance for it. It also means that the usual procedure of quantize the field in RS, which just restrict its domain of definition, is not correct, because this kind of restriction is not enough to have a well posed theory. If one would like to study the spinor field in RS and work with physical states and modes, the two possible ways are to construct suitable wave packets made as combinations of the modes $\Psi_{i\mathcal{M}}$, or to consider from the beginning a constrained hamiltonian, different from $H_{R}$ which automatically assures that the condition (~\ref{eq:Condizione}) is satisfied. The crucial point is however that this condition prevents any relationship between the quantization in RS and that in MS, because it means that RS should be treated as an manifold on its own, and not as a submanifold of MS in some way related to it. Consequently, the necessity of boundary conditions on the field should manifest itself also in the analysis of the Unruh effect, which concerns just this relationship between RS and MS quantum constructions. To show this will be our next problem.  
\section{Analysis of the Unruh effect}
In order to analyse this relationship, it is convenient to perform a quantization in Minkowski spacetime which is different from the standard one (that in plane waves) and which could be more easily compared with the Rindler one; namely, a quantization in terms of the Lorentz Momentum eigenfunctions defined in the whole Minkowski space, that represents analytical continuation of the $\Psi_{i,\mathcal{M}}$ we found before. The determination of these functions will require some preliminary steps.   
\subsection{Solutions in the other sectors and their relationship}
First of all we will study the relations between the solutions of Dirac equation in the different sectors F, L, P (see ~\ref{sec:Rind}).
Of course, we omit the passages and just write down the form of Dirac equation in the sector and its possible solutions . That is the following.
\begin{center}
{\large F sector:}
\end{center}
\be  \left( i\,\gamma^{0}\,\partial_{\rho}\,+\,\frac{i}{\rho}\,\gamma^{1}\,\partial_{\eta}\,+\,\frac{i}{2}\,\frac{1}{\rho}\,\gamma^{0}\,+\,i\,\gamma^{2}\,\partial_{y}\,+\,i\,\gamma^{3}\,\partial_{z}\,-\,m\right)\,\Psi\,=\,0 \ee

\be \Psi_{i\mathcal{M}}^{F,\,\pm}\,=\,M^{\pm}_{i}\left( X_{i}^{F}\,K_{i\mathcal{M}-\frac{1}{2}}(\pm\,i\,\kappa\,\rho)\,+\,Y_{i}^{F}\,K_{i\mathcal{M}+\frac{1}{2}}(\pm\,i\,\kappa\,\rho)\right)\,e^{-\,i\,\mathcal{M}\,\eta}\,e^{i\,k_{2}\,y\,+\,i\,k_{3}\,z} \ee

with $i=1,2$ and
\be X_{1}^{F}\,=\,\left( \begin{array}{c} k_{3} \\ i\,\left(k_{2}\,+\,i\,m \right) \\ i\,\left(k_{2}\,+\,i\,m \right) \\ k_{3} \end{array} \right) \;\;\;\;\;\;\;\;\;\;\;\;\;\;Y_{1}^{F}\,=\,\left( \begin{array}{c} 0 \\ \mp\,\kappa \\ \pm\,\kappa \\ 0 \end{array} \right) \ee
\be X_{2}^{F}\,=\,\left( \begin{array}{c} 0 \\ \pm\,\kappa \\ \mp\,\kappa \\ 0 \end{array} \right) \;\;\;\;\;\;\;\;\;\;\;\;\;\;Y_{2}^{F}\,=\,\left( \begin{array}{c} k_{3} \\ i\,\left(k_{2}\,-\,i\,m \right) \\ -\,i\,\left(k_{2}\,-\,i\,m \right) \\ -\,k_{3} \end{array} \right) \ee

\begin{center}
{\large L sector:}
\end{center}
\be \left(  i\,\frac{1}{\rho}\,\gamma^{0}\,\partial_{\eta}\,+\,i\,\gamma^{1}\,\partial_{\rho}\,-\,i\,\gamma^{2}\,\partial_{y}\,-\,i\,\gamma^{3}\,\partial_{z}\,+\,\frac{1}{2}\,i\,\frac{1}{\rho}\,\gamma^{1}\,+\,m \right) \Psi\,=\,0 \ee
\be \Psi_{i\mathcal{M}}^{L,\,\pm}\,=\,O_{i}^{\pm}\,\left( X_{i}^{L}\,K_{i\,\mathcal{M}-\frac{1}{2}}(\kappa\rho)\,+\,Y_{i}^{L}\,K_{i\,\mathcal{M}+\frac{1}{2}}(\kappa\rho) \right)\,e^{-\,i\,\mathcal{M}\,\eta}\,e^{i\,k_{2}\,y\,+\,i\,k_{3}\,z} \ee
with
\be X_{1}^{L}\,=\,\left( \begin{array}{c} k_{3} \\ i\,\left(k_{2}\,+\,i\,m \right) \\ i\,\left(k_{2}\,+\,i\,m \right) \\ k_{3} \end{array} \right) \;\;\;\;\;\;\;\;\;\;\;\;\;\;Y_{1}^{L}\,=\,\left( \begin{array}{c} 0 \\ -\,i\,\kappa \\ i\,\kappa \\ 0 \end{array} \right) \ee
\be X_{2}^{L}\,=\,\left( \begin{array}{c} 0 \\ -\,i\,\kappa \\ -\,i\,\kappa \\ 0 \end{array} \right) \;\;\;\;\;\;\;\;\;\;\;\;\;\; Y_{2}^{L}\,=\,\left( \begin{array}{c} k_{3} \\ i\,\left(k_{2}\,-\,i\,m \right) \\ -\,i\,\left(k_{2}\,-\,i\,m \right) \\ -\,k_{3} \end{array} \right)  \ee

\begin{center}
{\large P sector:}
\end{center}
\be  \left( i\,\gamma^{0}\,\partial_{\rho}\,+\,\frac{i}{\rho}\,\gamma^{1}\,\partial_{\eta}\,+\,\frac{i}{2}\,\frac{1}{\rho}\,\gamma^{0}\,-\,i\,\gamma^{2}\,\partial_{y}\,-\,i\,\gamma^{3}\,\partial_{z}\,+\,m\right)\,\Psi\,=\,0 \ee

\be \Psi_{i\mathcal{M}}^{P,\,\pm}\,=\,P_{i}^{\pm}\left( X_{i}^{P}\,K_{i\mathcal{M}-\frac{1}{2}}(\pm\,i\,\kappa\,\rho)\,+\,Y_{i}^{P}\,K_{i\mathcal{M}+\frac{1}{2}}(\pm\,i\,\kappa\,\rho)\right)\,e^{-\,i\,\mathcal{M}\,\eta}\,e^{i\,k_{2}\,y\,+\,i\,k_{3}\,z} \ee

with
\be X_{1}^{P}\,=\,\left( \begin{array}{c} k_{3} \\ i\,\left(k_{2}\,+\,i\,m \right) \\ i\,\left(k_{2}\,+\,i\,m \right) \\ k_{3} \end{array} \right) \;\;\;\;\;\;\;\;\;\;\;\;\;\;Y_{1}^{P}\,=\,\left( \begin{array}{c} 0 \\ \pm\,\kappa \\ \mp\,\kappa \\ 0 \end{array} \right) \ee
\be X_{2}^{P}\,=\,\left( \begin{array}{c} 0 \\ \mp\,\kappa \\ \mp\,\kappa \\ 0 \end{array} \right) \;\;\;\;\;\;\;\;\;\;\;\;\;\;Y_{2}^{P}\,=\,\left( \begin{array}{c} k_{3} \\ i\,\left(k_{2}\,-\,i\,m \right) \\ -\,i\,\left(k_{2}\,-\,i\,m \right) \\ -\,k_{3} \end{array} \right) \ee

As regards to the relationship between these solutions, this could be found by analytically continuing the functions across the event horizons, which represent branch points for these functions. Using the variables
\be x_{+}\,=\,x\,+\,t \;\;\;\;\;\,\;\;\,\;\;x_{-}\,=\,t\,-\,x \ee
the passages trough the horizons is given by the substitutions: $-x_{-}\rightarrow x_{-}e^{\pm i\pi}$ ($R\rightarrow F$) $x_{+}\rightarrow -x_{+}e^{\pm i\pi}$ ($F\rightarrow L$), $x_{-}\rightarrow-x_{-}e^{\pm i\pi}$ ($L\rightarrow P$), $-x_{+}\rightarrow x_{+}e^{\pm i\pi}$ ($P\rightarrow R$).
The result is that the solutions of the Dirac equation are linked by two possible paths of analytical continuation, namely
\begin{itemize}
\item that one corresponding (we name it A) to the transformation $-x_{\pm}\rightarrow x_{\pm}e^{+ i\pi}$ 
and linking in succession the functions
\be \Psi_{i,\mathcal{M}}^{R}\,\rightarrow\,\Psi_{i,\mathcal{M}}^{F,\,+}\,\rightarrow\,\Psi_{i,\mathcal{M}}^{L,\,+}\,\rightarrow\,\Psi_{i,\mathcal{M}}^{P,\,-}\,\rightarrow\,\Psi_{i,\mathcal{M}}^{R} \ee
\item that one corresponding to the transformation $-x_{\pm}\rightarrow x_{\pm}e^{- i\pi}$ (we name it B)
and linking in succession the functions
\be \Psi_{i,\mathcal{M}}^{R}\,\rightarrow\,\Psi_{i,\mathcal{M}}^{F,\,-}\,\rightarrow\,\Psi_{i,\mathcal{M}}^{L,\,-}\,\rightarrow\,\Psi_{i,\mathcal{M}}^{P,\,+}\,\rightarrow\,\Psi_{i,\mathcal{M}}^{R} \ee
\end{itemize}
Moreover, it is possible to demonstrate that the normalization factors of the different solutions are related each one other in such a way that, once determined $N$, all the others are determined consequently.
\subsection{Lorentz Momentum eigenfunctions in MS}
We now turn to the problem of finding a unified representation for the eigenfunctions of the Boost Generator, which holds true in the whole MS and which is everywhere analytical. 
Let's consider the integral representations of the Bessel functions $K_{\nu}(\rho)$ given by:
\bea K_{\nu}(\rho)\,=\,\frac{1}{2}\,e^{-\,\frac{i\pi\nu}{2}}\,\int_{-\infty}^{+\infty}e^{i\,\rho\,\sinh\vartheta}\,e^{\nu\,\vartheta}\,d\vartheta \label{eq:rappr1}\\ K_{\nu}(\rho)\,=\,\frac{1}{2}\,e^{\frac{i\pi\nu}{2}}\,\int_{-\infty}^{+\infty}e^{i\,\rho\,\sinh\vartheta}\,e^{-\,\nu\,\vartheta}\,d\vartheta \eea
Using the second one to express in an integral form and in minkowski coordinates the functions $K_{\nu}(\rho(t,x))e^{-\nu\eta(t,x)}$ (using the coordinate transformation ~\ref{eq:cooRin}), we obtain:
\bea \lefteqn{K_{\nu}(\kappa\rho)\,e^{-\,\nu\,\eta}\,=} \\
 &=& \frac{1}{2}\,e^{\frac{i\pi\nu}{2}}\,\int_{-\infty}^{+\infty}e^{i\,\kappa\left[ x\,\sinh\vartheta\,-\,t\,\cosh\vartheta\right]}\,e^{-\,\nu\,\vartheta}\,d\vartheta\,= \label{eq:rap1}\\ &=& \frac{1}{2}\,e^{\frac{i\pi\nu}{2}}\,\int_{-\infty}^{+\infty}\,P_{\vartheta}^{-}(t\,,\,x)\,e^{-\,\nu\,\vartheta}\,d\vartheta \eea
where $P^{-}_{\vartheta}(t,x)$ represent (2-dim) positive frequency plane waves with $\omega=\kappa\cosh\vartheta$ and $k_{x}=\kappa\sinh\vartheta$.
Using the first one, and with the same procedure, we have:
\bea K_{\nu}(\kappa\rho)\,e^{-\,\nu\,\eta}\,=\,\frac{1}{2}\,e^{-\,\frac{i\pi\nu}{2}}\,\int_{-\infty}^{+\infty}e^{i\,\kappa\left[ x\,\sinh\vartheta\,+\,t\,\cosh\vartheta\right]}\,e^{\nu\,\vartheta}\,d\vartheta\,= \label{eq:rap2} \\ = \,\frac{1}{2}\,e^{-\,\frac{i\pi\nu}{2}}\,\int_{-\infty}^{+\infty}\,P_{\vartheta}^{+}(t\,,\,x)\,e^{\nu\,\vartheta}\,d\vartheta \eea
where now $P_{\vartheta}^{+}(t\,,\,x)$ are (2-dim) plane waves with negative frequency $-\omega=-\kappa\cosh\vartheta$. What we found means that the functions $K_{\nu}(\rho(t,x))e^{-\nu\eta(t,x)}$ can be expressed equivalently as linear (integral) combination of positive or negative two dimensional plane waves.

Inserting these formulas into the expression for $\Psi_{i,\mathcal{M}}^{R}(t,x,y,z)$, not taking care of the normalization factor, gives 
\bea \Psi_{i\mathcal{M}}^{\mp}\left(t,x,y,\,z\right)\,=\,\frac{1}{2}N^{\mp}\left[ X_{i}^{R}\,e^{\pm\,\frac{i\pi}{2}\left(i\mathcal{M}-\frac{1}{2}\right)}\,\int_{-\infty}^{+\infty}e^{i\,\kappa\left[ x\,\sinh\vartheta\,\mp\,t\,\cosh\vartheta\right]}\,e^{\mp\left(i\mathcal{M}-\frac{1}{2}\right)\vartheta}\,d\vartheta\,+\right. \nonumber \\ +\left.Y_{i}^{R}\,e^{\pm\,\frac{i\pi}{2}\left(i\mathcal{M}+\frac{1}{2}\right)}\,\int_{-\infty}^{+\infty}e^{i\,\kappa\left[ x\,\sinh\vartheta\,\mp\,t\,\cosh\vartheta\right]}\,e^{\mp\left(i\mathcal{M}+\frac{1}{2}\right)\vartheta}\,d\vartheta \right]\times \nonumber \\ \times\,e^{i\,k_{2}\,y\,+\,i\,k_{3}\,z} \;\;\;\;\;\;\;\; \label{eq:minkbesmodsp}\eea
These are the global functions we were looking for. For them, the following properties hold true:
\begin{itemize}
\item they are well behaved (analytical) on the entire Minkowski manifold, except for the origin;
\item they are solutions of the Dirac equation;
\item they are eigenfunctions of the boost generator operator $M_{01}$, with eigenvalue $\mathcal{M}$;
\item they reduce themselves to the correct solutions of the Dirac equations in the different sectors;
\item they correspond each to one of the two possible paths of analytical continuation we mentioned before, namely $\Psi_{i\mathcal{M}}^{-}$ corresponds to the path A, and $\Psi_{i\mathcal{M}}^{+}$ corresponds to the path B, so we could say that the reason for the existence of two different global representation is the existence of two different paths of analytical continuation across the horizons;
\item they are orthonormalized with respect to the ordinary scalar product in MS, with normalization factors given by:
\be N^{\mp}=\frac{e^{\pm\frac{1}{2}\pi\mathcal{M}}}{2\pi\sqrt{\kappa}} \ee 
\end{itemize} 
We note also that these functions represent the analogous in the spinor case of the Gerlach's Minkowski Bessel modes for the scalar field ~\cite{Gerlach}
\subsection{Alternative quantization in MS}
Having obtained these global functions, we can perform the quantization of the spinor field in terms of them. We remember that the usual plane wave expansion is given by:
\be \Psi\,=\,\sum_{r\,=\,1,2}\int d^{3}k\,\left[a_{r}(k)\,\Psi_{r}^{+}(k)\,+\,b_{r}^{\dagger}(k)\,\Psi_{r}^{-}(k)\right] \label{eq:esppian} \ee  
where the $\Psi_{r}^{+}(k)$ are positive frequency plane waves and $\Psi_{r}^{-}(k)$ are negative frequency ones, and that the quantum vacuum state is defined by the relation:
\be a_{r}(k)\,\mid 0\,\rangle_{M}\,=\,b_{r}(k)\,\mid 0\,\rangle_{M}\,=\,0 \;\;\;\;\,\;\;\;\forall\,r\,,\,\vec{k} \ee

But we knows that our $\Psi_{i\mathcal{M}}^{-}$ are linear combinations of positive frequency plane waves and  $\Psi_{i\mathcal{M}}^{+}$ of negative frequency ones, so we can have this kind of expansion:
\be \Psi(t,x,y,z)\,=\sum_{i\,=\,1,2}\int_{-\infty}^{+\infty} d\mathcal{M}\int dk_{2}\int dk_{3}\,\left[ c_{i\mathcal{M}}(k_{2},k_{3})\,\Psi_{i\mathcal{M}}^{-}\,+\,d^{\dagger}_{i\mathcal{M}}(k_{2},k_{3})\,\Psi_{i\mathcal{M}}^{+}\right] \label{eq:esplor} \ee 
then imposing the usual anticommutations rules
\be \left\{ c_{i\mathcal{M}}(k_{2},k_{3})\,,\,c^{\dagger}_{j\mathcal{M}'}(k'_{2},k'_{3}) \right\} \,=\,\delta_{ij}\,\delta\left( \mathcal{M}-\mathcal{M}'\right)\,\delta\left( k_{2}\,-\,k_{2}'\right)\delta\left( k_{3}\,-\,k_{3}'\right) \ldots\ldots \label{eq:regcomm}\ee
so defining a vacuum state $\mid 0 \rangle$ by means of:
\be c_{i\mathcal{M}}(k_{2},k_{3})\,\mid 0\,\rangle\,=\,d_{i\mathcal{M}}(k_{2},k_{3})\,\mid 0\,\rangle\,=\,0\,\;\;\,\;\;\;\,\;\;\;\forall\,i\,,\,\mathcal{M}\,,\,k_{2}\,,\,k_{3} \ee
It is easy now to show that this quantization in equivalent to the usual one and so that the state $\mid 0 \rangle$ is the usual Minkowski vacuum state $\mid 0\rangle_{M}$.
In fact we have the following relations:
\bea \lefteqn{c_{i\mathcal{M}}(k_{2},k_{3})\,=\,\left( \Psi_{i\mathcal{M}}^{-}\,,\,\Psi\right)_{M}\,=\,\int_{M}d^{3}x\,\Psi_{i\mathcal{M}}^{-\,\dagger}\,\Psi\,=} \nonumber \\ &=& \sum_{r}\,\int_{0}^{+\infty}dk'_{1}\,\frac{2\,\pi^{2}}{\omega'}\,N^{-}\,N_{k'}\,\left[  e^{i\,\frac{\pi}{2}\left( i\mathcal{M}+\frac{1}{2}\right)}\,e^{\left( i\mathcal{M}+\frac{1}{2}\right)\vartheta'}\,X_{i}^{\dagger}u_{r}(k')\,+ \nonumber \right. \\ &+&\left.\,e^{i\,\frac{\pi}{2}\left( i\mathcal{M}-\frac{1}{2}\right)}\,e^{\left( i\mathcal{M}-\frac{1}{2}\right)\vartheta'}\,Y_{i}^{\dagger}u_{r}(k')\right]\,\times\,a_{r}\left(k'_{1},k_{2},k_{3}\right)\,= \label{eq:c}\\ &=&\,\sum_{r}\,\int_{0}^{+\infty}dk'_{1}\,F_{i\,r}\left( k'_{1}\,,\,\mathcal{M}\right)\,a_{r}\left(k'_{1},k_{2},k_{3}\right) \eea
with $\vartheta'=\frac{1}{2}\,\ln\left(\frac{\omega'\,+\,k'_{1}}{\omega'\,-\,k'_{1}}\right)$

and, for $d_{i\mathcal{M}}^{\dagger}(k_{2},k_{3})$:
\bea \lefteqn{d_{i\mathcal{M}}^{\dagger}(k_{2},k_{3})\,=\,\left( \Psi_{i\mathcal{M}}^{+}\,,\,\Psi\right)\,=\,\int_{M}d^{3}x\,\Psi_{i\mathcal{M}}^{+\,\dagger}\,\Psi\,=} \nonumber \\ &=& \sum_{r}\,\int_{0}^{+\infty}dk'_{1}\,\frac{2\,\pi^{2}}{\omega'}\,N^{+}\,N_{k'}\,\left[  e^{i\,\frac{\pi}{2}\left( i\mathcal{M}+\frac{1}{2}\right)}\,e^{\left( i\mathcal{M}+\frac{1}{2}\right)\vartheta'}\,X_{i}^{\dagger}v_{r}(k')\,+ \right. \nonumber \\ &+& \left.\,e^{i\,\frac{\pi}{2}\left( i\mathcal{M}-\frac{1}{2}\right)}\,e^{\left( i\mathcal{M}-\frac{1}{2}\right)\vartheta'}\,Y_{i}^{\dagger}v_{r}(k')\right]\,\times\,b^{\dagger}_{r}\left(k'_{1},k_{2},k_{3}\right)\,= \\ &=&\,\sum_{r}\,\int_{0}^{+\infty}dk'_{1}\,G_{i\,r}\left( k'_{1}\,,\,\mathcal{M}\right)\,b^{\dagger}_{r}\left(k'_{1},k_{2},k_{3}\right) \eea

So it is demonstrated that the vacuum states defined by the two quantization procedures are the same. Moreover, by explicit calculation it is possible to show that
\bea \lefteqn{\left\{ a_{s}\left( k''_{1},k_{2},k_{3}\right)\,,\,a_{r}^{\dagger}\left( k'_{1},k'_{2},k'_{3}\right)\right\}\,=\,\delta_{rs}\,\delta\left(k'_{1}-k''_{1}\right)\,\delta\left( k_{2}-k'_{2}\right)\,\delta\left(k_{3}-k'_{3}\right)\;\;\;\Longleftrightarrow} \nonumber \\ &\Longleftrightarrow&\;\left\{ c_{i\mathcal{M}}(k_{2},k_{3})\,,\,c^{\dagger}_{j\mathcal{M}'}(k'_{2},k'_{3})\right\}\,=\,Cost\,\times\,\delta_{ij}\,\delta\left(\mathcal{M}-\mathcal{M}'\right)\delta\left( k_{2}-k'_{2}\right)\,\delta\left(k_{3}-k'_{3}\right) \nonumber \eea
so the two quantum constructions are totally equivalent. This means also that the divergence of the boost modes at the origin of MS does not create any qualitative difficulty for the quantization procedure, but just requires some additional care in the calculations we will perform in the following.
\subsection{The Unruh construction and the Unruh effect} \label{sec:Un}
We will now derive the Unruh effect for the spinor field following the standard procedure first used by Unruh himself. The main idea is to have a quantum construction which is valid for MS and reproduces the construction outlined in ~\ref{sec:RQFT} for the R sector. Let's perform the Unruh construction for quantization in MS, using our globally defined functions $\Psi_{i\mathcal{M}}^{\mp}$.
Consider the functions:
\bea \lefteqn{R_{i\mathcal{M}}\,=\,\frac{1}{\sqrt{2\,\cosh\pi\mathcal{M}}}\,\left( e^{\frac{\pi\,\mathcal{M}}{2}}\,\Psi_{i\mathcal{M}}^{-}\,+\,e^{-\,\frac{\pi\,\mathcal{M}}{2}}\,\Psi_{i\mathcal{M}}^{+}\right)}\label{eq:RU} \\ &L_{i\mathcal{M}}&\,=\,\frac{1}{\sqrt{2\,\cosh\pi\mathcal{M}}}\,\left( e^{-\,\frac{\pi\,\mathcal{M}}{2}}\,\Psi_{i\mathcal{M}}^{-}\,-\,e^{\frac{\pi\,\mathcal{M}}{2}}\,\Psi_{i\mathcal{M}}^{+}\right) \label{eq:LU}\eea
which are solutions of Dirac equations, are eigenfunctions of $M_{01}$, are analytical in the whole Minkowski space, and orthonormalzed in MS. Moreover, it happens that the $R_{i\mathcal{M}}$ are defined everywhere but in L sector and reduce themselves to the $\Psi_{i\mathcal{M}}^{R}$ in the R one, while the $L_{i\mathcal{M}}$ manifest the inverse behaviour. Inverting these relations, and inserting into the expansion (~\ref{eq:esplor}), we have:
\bea \lefteqn{\Psi(t,x,y,z)\,=\sum_{i\,=\,1,2}\int_{-\infty}^{+\infty} d\mathcal{M}\int dk_{2}\int dk_{3}\,\left[ c_{i,\mathcal{M}}(k_{2},k_{3})\,\Psi_{i,\mathcal{M}}^{-}\,+\,d^{\dagger}_{i,\mathcal{M}}(k_{2},k_{3})\,\Psi_{i,\mathcal{M}}^{+}\right]\,=} \nonumber \\  &=& \sum_{i}\int_{-\infty}^{+\infty} d\mathcal{M}\left[ r_{i,\mathcal{M}}\,R_{i,\mathcal{M}}\,+\,l^{\dagger}_{i,\mathcal{M}}\,L_{i,\mathcal{M}}\right]\,= \label{eq:espUnr}\\ &=& \sum_{i}\int_{0}^{+\infty} d\mathcal{M}\left[ r_{i,\mathcal{M}}\,R_{i,\mathcal{M}}\,+\,l_{i,\mathcal{M}}\,L_{i,-\mathcal{M}}\,+\,r^{\dagger}_{i,\mathcal{M}}\,R_{i,-\mathcal{M}}\,+\,l^{\dagger}_{i,\mathcal{M}}\,L_{i,\mathcal{M}}\right]\;\;\;\;\;\,\;\;\;\;\;\; \eea
having introduced the coefficients:
\bea r_{i,\mathcal{M}}\,=\,\frac{c_{i,\mathcal{M}}\,e^{\frac{\pi\mathcal{M}}{2}}\,+\,d^{\dagger}_{i,\mathcal{M}}\,e^{-\frac{\pi\mathcal{M}}{2}}}{\sqrt{2\,\cosh\pi\mathcal{M}}} \label{eq:Bog} \\ l^{\dagger}_{i,\mathcal{M}}\,=\,\frac{c_{i,\mathcal{M}}\,e^{-\frac{\pi\mathcal{M}}{2}}\,-\,d^{\dagger}_{i,\mathcal{M}}\,e^{\frac{\pi\mathcal{M}}{2}}}{\sqrt{2\,\cosh\pi\mathcal{M}}} \eea
These relations represent the Bogolubov transformation between the quantum constructions (~\ref{eq:esplor}) and (~\ref{eq:espUnr}).

Let's now consider the operator $N_{i,j,\mathcal{M},\mathcal{M}'}=r_{i\mathcal{M}}^{\dagger}r_{j\mathcal{M}'}$ (we can of course not consider the quantum numbers $k_{2}$ and $k_{3}$, because they don't play a significant role here). We remind that $\mathcal{M}$ is the eigenvalue of $M_{01}$ but also the energy of a Rindler observer. It is easy to calculate the mean value of this operator in the Minkowski vacuum state, having:
\bea \lefteqn{_{M}\langle 0\,\mid\,N_{i,j,\mathcal{M},\mathcal{M}'}\,\mid 0\,\rangle_{M}\,=\,_{M}\langle 0\,\mid\,r_{j\mathcal{M}'}^{\dagger}r_{i\mathcal{M}}\,\mid 0\,\rangle_{M}\,=} \\  &=& \frac{1}{e^{2\pi\mathcal{M}}\,+\,1}\,\delta_{ij}\,\delta\left(\mathcal{M}-\mathcal{M}'\right)\,\delta\left( k_{2}-k'_{2}\right)\delta\left( k_{3}-k'_{3}\right) \label{eq:numpart}\eea
If we now identify the operators $r_{i,\mathcal{M}}$ with the operators $a_{i,\mathcal{M}}$ (Rindler annihilation operators), then we can intepret $N_{i,j,\mathcal{M},\mathcal{M}'}$ as a Rindler particle number operator. This is exactly what is usually done in literature. The reasons for this identification is the particular functional behaviour of the $R_{i\mathcal{M}}$, that we have just mentioned. Nevertheless this passage is not trivial at all, as we will show.

Anyway, once identified the operators $r_{i,\mathcal{M}}$ with the operators $a_{i,\mathcal{M}}$, the result (~\ref{eq:numpart}) can be interpreted as meaning that an inertial observer and a Rindler observer don't share the same vacuum state, and that Minkowski vacuum state is a particle state for a Rindler observer.

We can also calculate the total number of particles that a Rindler observer will perceive if the field is in Minkowski vacuum state, for any quantum number and for unity of proper time.
The result is:
\bea \lefteqn{dN\,=\,\sum_{i,j}\int_{0}^{+\infty}d\mathcal{M}\int_{0}^{+\infty}d\mathcal{M}'\int dk_{2}\int dk'_{2}\int dk_{3}\int dk'_{3}\,N_{i,j,\mathcal{M},\mathcal{M}'}\,d\eta\,=} \\ &=&\,\sum_{i,j}\int_{0}^{+\infty}d\mathcal{M}\int_{0}^{+\infty}d\mathcal{M}'\int dk_{2}\int dk'_{2}\int dk_{3}\int dk'_{3} \,_{M}\langle 0\,\mid\,r_{j\mathcal{M}'}^{\dagger}r_{i\mathcal{M}}\,\mid 0\,\rangle_{M}\,d\eta\,= \nonumber \\ &=& \sum_{i,j}\int_{0}^{+\infty}d\mathcal{M}\int_{0}^{+\infty}d\mathcal{M}'\int dk_{2}\int dk'_{2}\int dk_{3}\int dk'_{3}\,\frac{1}{e^{2\pi\mathcal{M}}\,+\,1}\,\times \\ &\times& \delta_{ij}\,\delta\left(\mathcal{M}-\mathcal{M}'\right)\,\delta\left( k_{2}-k'_{2}\right)\delta\left( k_{3}-k'_{3}\right)\,d\eta\,= \\ &=& \sum_{i}\int_{0}^{+\infty}d\mathcal{M}\int dk_{2}\int dk_{3}\,\frac{1}{e^{2\pi\mathcal{M}}\,+\,1}\,d\eta\,= \\ &=&\,\sum_{i}\int_{0}^{+\infty}dh_{R}\int dk_{2}\int dk_{3}\,\frac{1}{e^{\frac{2\pi\,h_{R}}{a}}\,+\,1}\,d\tau\,= \\ &=& 2\,\int_{0}^{+\infty}dh_{R}\int dk_{2}\int dk_{3}\,\frac{1}{e^{\frac{2\pi\,h_{R}}{a}}\,+\,1}\,d\tau   \label{eq:flussoU} \eea
where we have used the quantities: $h_{R}=a\mathcal{M}$, Rindler energy, and $\tau=\frac{\eta}{a}$, proper time, with $a$ acceleration of the Rindler observer.
This result could be stated saying that the particle distribution of the Minkowski vacuum state, with respect to the quantization performed using the $R_{i\mathcal{M}}$ modes, is given by a thermal spectrum, according to Fermi-Dirac statistics with temperature:
\be T\,=\,\frac{1}{\beta\,k_{B}}\,=\,\frac{a}{2\,\pi\,k_{B}} \label{eq:TTT}\ee
This is the Unruh effect.

We saw that the identification between the operators $r_{i,\mathcal{M}}$ and $a_{i,\mathcal{M}}$ is crucial in its interpretation.  
Consequently, we will now analyse in detail the relationship between the above operators, trying to understand if this interpretation is well confirmed.
\subsection{Analysis of the coefficients}
We now are going to find the explicit expression of the coefficients $r_{i\mathcal{M}}$ as functions of the values of the field $\Psi$.
We remind that these are defined as:
\be r_{i,\mathcal{M}}\,=\,\frac{c_{i,\mathcal{M}}\,e^{\frac{\pi\mathcal{M}}{2}}\,+\,d^{\dagger}_{i,\mathcal{M}}\,e^{-\frac{\pi\mathcal{M}}{2}}}{\sqrt{2\,\cosh\pi\mathcal{M}}} \label{eq:r}\ee 
so we first need to find the expression for the coefficients $c_{i,\mathcal{M}}$  and $d^{\dagger}_{i,\mathcal{M}}$, in terms of the field and of its spatial derivative. This task implies the proper treatment of integral whose hypersurface of integration is the $t=0$ hypersurface, which pass across the origin of Minkowski space, that is the intersection of the branch points where the functions we used are not well defined; this requires great attention.

By performing this calculation it is also possible to see that, as it should be, the coefficients $c_{i,\mathcal{M}}$  and $d^{\dagger}_{i,\mathcal{M}}$ are well defined without the need of any additional boundary condition other than the vanishing of the field at spatial infinity, in contrast to the \lq\lq Rindler coefficients" in (~\ref{eq:exp}).

Inserting the expressions so found for $c_{i,\mathcal{M}}$  and $d^{\dagger}_{i,\mathcal{M}}$ into the ~\ref{eq:r}, we obtain:
\bea \lefteqn{r_{i,\mathcal{M}}\,=} \nonumber \\ &=& \frac{1}{2\,\pi\,\sqrt{\kappa}}\,\sqrt{2\,\cosh\pi\mathcal{M}}\,\int_{-\infty}^{+\infty}dy\,e^{-ik_{2}y}\,\int_{-\infty}^{+\infty}dz\,e^{-ik_{3}z}\,\times \nonumber \\ &\times& \left\{ X_{i}^{\dagger}\,\int_{0}^{+\infty}dx\,\left[ K_{i\mathcal{M}+\frac{1}{2}}(\kappa x)\Psi\left( 0\,,\,x\right)\,-\right.\right. \nonumber \\ &-&\,\frac{1}{2}\,\Gamma\left( \frac{1}{2}+ i\mathcal{M}\right)\left(\frac{\kappa x}{2}\right)^{- i\mathcal{M}-\frac{1}{2}}\,\Psi\left(0\,,\,x\right)\,- \nonumber \\ &-&\left.\,\frac{1}{\kappa}\,\frac{1}{\frac{1}{2}- i\mathcal{M}}\,\Gamma\left( \frac{1}{2}+ i\mathcal{M}\right)\left(\frac{\kappa x}{2}\right)^{- i\mathcal{M}+\frac{1}{2}}\,\frac{d}{dx}\Psi\left(0\,,\,x\right)\right]\,+ \nonumber \\ &+& Y_{i}^{\dagger}\,\int_{0}^{+\infty}dx\,\left[ K_{i\mathcal{M}-\frac{1}{2}}(\kappa x)\Psi\left( 0\,,\,x\right)\,-\right. \nonumber \\ &-&\,\frac{1}{2}\,\Gamma\left( \frac{1}{2}- i\mathcal{M}\right)\left(\frac{\kappa x}{2}\right)^{+ i\mathcal{M}-\frac{1}{2}}\,\Psi\left(0\,,\,x\right)\,- \nonumber \\ &-&\left.\left.\,\frac{1}{\kappa}\,\frac{1}{\frac{1}{2}+ i\mathcal{M}}\,\Gamma\left( \frac{1}{2}- i\mathcal{M}\right)\left(\frac{\kappa x}{2}\right)^{+ i\mathcal{M}+\frac{1}{2}}\,\frac{d}{dx}\Psi\left(0\,,\,x\right)\right] \right\} \label{eq:r1}\eea
Changing coordinates to the Rindler ones, we have:
\bea \lefteqn{r_{i,\mathcal{M}}\,=} \nonumber \\ &=& \sqrt{2}\,a_{i,\mathcal{M}}\,+\,\frac{1}{2\,\pi\,\sqrt{\kappa}}\,\sqrt{2\,\cosh\pi\mathcal{M}}\,\int_{-\infty}^{+\infty}dy\,e^{-ik_{2}y}\,\int_{-\infty}^{+\infty}dz\,e^{-ik_{3}z}\,\times \nonumber \\ &\times& \left\{ \lim_{\rho\rightarrow 0}\,\frac{1}{2}\,\frac{1}{\frac{1}{2}- i\mathcal{M}}\,\Gamma\left( \frac{1}{2}+ i\mathcal{M}\right)\left(\frac{\kappa \rho}{2}\right)^{- i\mathcal{M}+\frac{1}{2}}\,X_{i}^{\dagger}\,\Psi\left(0\,,\,\rho\right)\,+\right. \nonumber \\ &+&\,\left.\lim_{\rho\rightarrow 0}\,\frac{1}{2}\,\frac{1}{\frac{1}{2}+ i\mathcal{M}}\,\Gamma\left( \frac{1}{2}- i\mathcal{M}\right)\left(\frac{\kappa \rho}{2}\right)^{+ i\mathcal{M}+\frac{1}{2}}\,Y_{i}^{\dagger}\,\Psi\left(0\,,\,\rho\right)\right\} \eea

It is so clear that the coefficients $r_{i,\mathcal{M}}$ and $a_{i,\mathcal{M}}$ cannot be identified unless
\be \lim_{\rho\,\rightarrow\,0}\,\rho^{\frac{1}{2}}\,\Psi\left( 0\,,\,\rho\right)\,=\,0 \label{eq:co} \ee
which is just the boundary condition we found in sec.~\ref{sec:RQFT}.

Let's now discuss this result.
\subsection{Discussion of the result}
We saw that it is not possible to identify the operators $r_{i,\mathcal{M}}$ and $a_{i,\mathcal{M}}$, unless (~\ref{eq:co}); but
there are no physical reasons to impose this boundary condition on the spinor field in MS;  
this problem could be avoided only if we perform the \lq\lq Unruh construction" just in the R and L sectors of MS, which are completely disjoint, from the causal and, if we impose (~\ref{eq:co}), physical point of view;
in other words, the Unruh construction outlined in \ref{sec:Un} does not represent a valid quantization scheme for the whole MS;
moreover, for an observer living in the Rindler wedge it is not possible, because of condition (~\ref{eq:co}), to perform measurement in the whole MS, in order to put the field in the Minkowski vacuum state;     
therefore, the relation (~\ref{eq:flussoU}) cannot be interpreted in any sense as a proof of any \lq\lq Unruh effect";
the conclusion can only be that there is no relation between the quantum field theory that could be contructed in RS and that one in MS, so to speak about thermal properties of the Minkowski vacuum state with respect to RS is meaningless.

\section{Conclusions}
Here we come to our conclusions. They could be summarized by the statement: the basic principles of quantum field theory give no grounds to assert that the Unruh effect exists.

We saw that the reason for this conclusion is the existence of boundary conditions for the quantization of the spinor field in Rindler spacetime, preventing any relationship between this quantum construction and that one in Minkowski spacetime. Moreover the necessity of these boundary conditions was analytically showed in our analysis of the Unruh's procedure to derive his effect. We already noted that the existence of such boundary conditions could be expected since a Rinlder observer is confined inside the Rindler wedge by event horizons, so he would see these like spatial infinity of an inertial observer.
For the same reason a Rindler observer has no relationship with MS and he cannot in any way prepare the quantum field in the Minkowski vacuum state.
 
It seems to us that there are grounds to assert that this is a general feature of the analysed problem and that it holds true for any quantum field, and we are supported in this conclusion by the previously mentioned and similar results obtained for the scalar field.

In ~\cite{Bel3} it is also well shown that the Unruh quantization scheme is valid only in the double Rindler wedge $R\cup L$, and so it is also in the spinor case, because of the boundary condition (~\ref{eq:co}); consequently the Unruh construction cannot be used to analyse the relationship between Rindler and Minkowski quantization schemes.    

On the other hand, the appearence of the fermion form in the distribution (~\ref{eq:flussoU}) is entirely due to the particular form of the Bogolubov transformation (~\ref{eq:Bog}), and there is no real need to interpretate it as an proof of a thermal nature of the spectrum; the same situation, again, was found in the case of the scalar field, and it is explained in details in ~\cite{Bel3}.

As regards to the other aspect of the Unruh problem, that we mentioned in the beginning, namely the behaviour of an accelerated detector in Minkowski space, we could only say that it remains an open question. This should be clear also just considering that the Unruh effect is generally explained using the key role of the event horizons, but there are no horizon at all for a non-ideal accelerating detector, whose acceleration lasts a finite amount of time. 

Our results show, however, that there are no reasons to exspect that it will be of the type predicted by Unruh, and that it should not be expected to be universal and independent from the nature and characteristics of the detector itself. It is worth to note that, as a partial confirmation of what we are saying, it was showed in great details in ~\cite{NikRit} that elementary particles accelerated by a constant electric field don't follow, in general, the Unruh behaviour, i.e. the thermal response with temperature (\ref{eq:T}).

We are sure, of course, that an accelerated detector behaves, in general, in a different way from an inertial one, and we admit that in some cases it could follow the Unruh behaviour, but we proved that there are no quantum theoretical reasons to believe that this is the universal one. 
\section*{Acknowledgements} 
We would like to thank here Prof. V. A. Belinski, for the enlightening discussions had with him during the year in which this work was done, for his permanent help and encouragement, and for having read the manuscript of this paper, giving useful comments. A special thank goes also to Prof. Narozhny, who noticed an error occurred in a previous version of this work, now corrected, and to the organizing committee of the Third ICRA Network Workshop, expecially Prof. R. Ruffini, which gave us the possibility to present our results in that occasion.

\end{document}